\documentclass[aps,pra,twocolumn,showpacs,preprintnumbers,amsmath,amssymb,superscriptaddress,10pt]{revtex4-1}
\usepackage{amsmath,amssymb}
\usepackage{bm}% bold math
\usepackage[latin1]{inputenc}
\usepackage{dcolumn}% Align table columns on decimal point
\usepackage{graphicx}% Include figure files
\usepackage{SIunits}
\usepackage{multirow}

\def\oscwidth{8.6cm}
\def\plotwidth{8cm}

\begin{document}
%\preprint{APS/123-QED}

\title{Thermal blinding of gated detectors in quantum cryptography}% Force line breaks with\\

\author{Lars Lydersen}
\email{lars.lydersen@iet.ntnu.no}
\affiliation{Department of Electronics and Telecommunications, Norwegian University of Science and Technology, NO-7491 Trondheim, Norway}
\affiliation{University Graduate Center, NO-2027 Kjeller, Norway}

\author{Carlos Wiechers}
\affiliation{Max Planck Institute for the Science of Light, G\"{u}nther-Scharowsky-Str. 1/Bau 24, 91058 Erlangen, Germany}
\affiliation{Institut f\"{u}r Optik, Information und Photonik, University of Erlangen-Nuremberg, Staudtstra\ss e 7/B2, 91058 Erlangen, Germany}
\affiliation{Departamento de F\'{i}sica, Universidad de Guanajuato, Lomas del Bosque 103, Fraccionamiento Lomas del Campestre, 37150, Le\'{o}n, Guanajuato, M\'{e}xico}

\author{Christoffer Wittmann}
\author{Dominique Elser}
\affiliation{Max Planck Institute for the Science of Light, G\"{u}nther-Scharowsky-Str. 1/Bau 24, 91058 Erlangen, Germany}
\affiliation{Institut f\"{u}r Optik, Information und Photonik, University of Erlangen-Nuremberg, Staudtstra\ss e 7/B2, 91058 Erlangen, Germany}

\author{Johannes Skaar}
\affiliation{Department of Electronics and Telecommunications, Norwegian University of Science and Technology, NO-7491 Trondheim, Norway}
\affiliation{University Graduate Center, NO-2027 Kjeller, Norway}

\author{Vadim Makarov}
\affiliation{Department of Electronics and Telecommunications, Norwegian University of Science and Technology, NO-7491 Trondheim, Norway}

\date{\today}% It is always \today, today,

\begin{abstract}
It has previously been shown that the gated detectors of two commercially available quantum key distribution (QKD) systems are blindable and controllable by an eavesdropper using continuous-wave illumination and short bright trigger pulses, manipulating voltages in the circuit [L{.} Lydersen \emph{et al{.}}, Nat{.} Photonics DOI:10.1038/nphoton.2010.214]. This allows for an attack eavesdropping the full raw and secret key without increasing the quantum bit error rate (QBER). Here we show how thermal effects in detectors under bright illumination can lead to the same outcome. We demonstrate that the detectors in a commercial QKD system Clavis2 can be blinded by heating the avalanche photo diodes (APDs) using bright illumination, so-called \emph{thermal blinding}. Further, the detectors can be triggered using short bright pulses once they are blind. For systems with pauses between packet transmission such as the plug-and-play systems, thermal inertia enables Eve to apply the bright blinding illumination \emph{before} eavesdropping, making her more difficult to catch.
\end{abstract}

\maketitle

\section{Introduction}
\label{sec:intro}
In theory quantum mechanics allows two parties, Alice and Bob, to grow a private, secret key, even if the eavesdropper Eve can do anything permitted by the laws of nature \cite{bennett1984,ekert1991,lo1999,shor2000}. The field of quantum key distribution (QKD) has evolved rapidly in the last two decades, with transmission distance increasing from a table top demonstration to over $250\,\kilo\meter$ in the laboratory \cite{stucki2009}, and commercial QKD systems available from several vendors\cite{comqkdsystems}.

However the components used for the experimental realizations of QKD have imperfections. Numerous imperfections have been addressed in security proofs \cite{mayers1996,gottesman2004,inamori2007,fung2009,lydersen2010,maroy2009}. For some loopholes it took several years from their discovery until they were covered by security proofs, for instance the Trojan-horse \cite{vakhitov2001,gisin2006} loophole and detector efficiency mismatch \cite{makarov2006,makarov2008}. The latter was exploited in the time-shift attack \cite{qi2007} on a commercial QKD system \cite{zhao2008}. Other loopholes include a variety of side-channels \cite{lamas-linares2007,nauerth2009,fung2007,xu2010}.

Common to the loopholes mentioned so far is that they are not implementable in practice, or only leave a marginal advantage for Eve. For instance, the implementation of the time-shift attack \cite{zhao2008} gave Eve an information-theoretic advantage, allowing her to outperform a straight brute-force search for the key in 4\% of her attempts. In the practical phase-remapping attack \cite{xu2010}, Eve caused 19.7\% QBER compromising merely the hardly ever used two-way post-processing protocol which produces secure key at QBER up to 20\% \cite{chau2002,gottesman2003}.

There is however one class of attacks which stands out in terms of implementability, Eve's information and QBER: The \emph{blinding attacks} \cite{makarov2009,makarov2008a,lydersen} are fully implementable with current technology, and give Eve the whole raw key while causing zero additional QBER. In these attacks, the APDs are tricked to exit the single-photon sensitive Geiger mode, and are so-called \emph{blind}. Eve uses a copy of Bob's apparatus to detect Alice's signals, but resends bright trigger pulses instead of single photons, as in the after-gate attack \cite{wiechers}. When the detectors are blind, Bob will only detect the bright trigger pulses if he uses the same basis as Eve. Otherwise his detectors remain silent. Hence Eve gets a full copy of the raw key while causing no additional QBER. Both passively quenched detectors \cite{makarov2009}, actively quenched detectors \cite{makarov2008a} and the gated detectors of two commercially available QKD systems \cite{lydersen} have been shown to be vulnerable to blinding. In the case of the passively-quenched detectors, this loophole has been exploited in the first full-scale implementation of an eavesdropper \cite{gerhardt}, which was inserted in the middle of the $290\,\meter$ transmission line in an experimental entanglement-based QKD system \cite{marcikic2006,peloso2009}, and recovered 100\% of the raw key.

Previously the gated detectors in the commercially available system Clavis2 from manufacturer ID Quantique were subject to continuous-wave (CW) blinding \cite{lydersen}. The blinding illumination caused the bias voltage at the APDs to drop due to the presence of DC impedance of the bias voltage supply, and therefore the APDs were never in Geiger mode. In this paper we show how the same detectors, regardless of the impedance of the bias voltage supply, can be blinded by heating the APD, so-called \emph{thermal blinding}. We show that thermal blinding is more sophisticated form of attack than previously reported CW-blinding \cite{lydersen} because the APD can be heated well in advance of the detection times, and is as such harder to catch. Especially for Clavis2, all the detector parameters such as temperature of the cold plate, bias voltage and APD current indicate single photon sensitivity while the detectors are in fact blind.

In this paper we first briefly review how APDs in the linear mode can be exploited to eavesdrop on QKD systems (section~\ref{sec:attack}). Then the detector design in Clavis2 is discussed (section~\ref{sec:detectors}) before we show how it is possible to thermally blind and trigger the detectors (section~\ref{sec:blinding}). Finally we briefly discuss countermeasures in section~\ref{sec:countermeasures} and conclude in section~\ref{sec:conclusion}.

\section{Eavesdropping exploiting APD\MakeLowercase{s} in linear mode}
\label{sec:attack}
In this section we briefly review how APDs in the linear mode can be exploited to eavesdrop on QKD systems \cite{makarov2008a,lydersen}.

In Geiger mode operation, an electron-hole pair produced by an absorbed single photon is amplified to a large current in the APD, which exceeds a current comparator threshold and reveals the photon's presence. This is referred to as a \emph{click} \cite{cova2004}.

In the linear mode however, when an APD is reverse-biased at a constant voltage below the breakdown voltage \footnote{All references to the APD bias voltage are absolute valued, thus an APD biased ``above'' the breakdown voltage is in the Geiger mode. In practice the APDs are always reverse-biased.}, the current through the APD is proportional to the incident optical power. Usually the APD is placed in a resistive network, and also has an internal resistance. Hence, the current through the APD lowers the bias voltage, and the current through the APD is monotonically increasing with the incident optical power. In this regime, the comparator current threshold translates to a classical optical power threshold \cite{lydersen}.

\begin{figure}[htbp]
  \includegraphics[width=8cm]{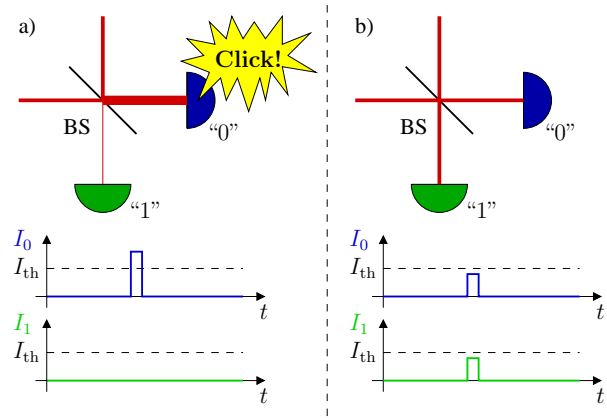}
  \caption{The last beam splitter (BS) as well as the detectors in a phase-encoded QKD system. $I_0$ and $I_1$ is the current running through APD 0/1, and $I_{th}$ is the comparator threshold current above which the detector registers a click. Here we assume that the APDs are in the linear mode, and that Eve sends a bright pulse slightly above the optical power thresholds. a) Eve and Bob have selected matching bases. Therefore the full intensity in the pulse from Eve hits detector~0. The current caused by Eve's pulse crosses the threshold current and causes a click. b) Eve and Bob have selected opposite bases. Therefore half the intensity of Eve's pulse hits each detector (corresponding to 50\% detection probability in either detector for single photons). This causes no click as the current is below the threshold for each detector.}
  \label{fig:attack} 
\end{figure}

If APDs are used as detectors in a QKD system, and they are optically accessible to Eve when biased under the breakdown voltage, Eve may eavesdrop on the QKD system with an intercept-resend (faked-state \cite{makarov2005}) attack. Eve uses a copy of Bob to detect the qubits from Alice in a random basis. Eve resends her detection results, but instead of sending single photons she sends bright pulses, just above the classical optical power threshold. Bob will only have a detection event if his basis choice coincides with Eve's basis choice (see Fig.~\ref{fig:attack}), otherwise no detector clicks.

After the raw key exchange, Bob and Eve are identical both in bit values and basis choices. Since Eve uses a copy of Bob's detectors, Bob's photon-number detection statistics is equal with or without Eve. Therefore the attack works equally well on the BB84 protocol \cite{bennett1984}, the Scarani-Acin-Ribordy-Gisin 2004 (SARG04) \cite{scarani2004} and decoy-state BB84 protocols \cite{hwang2003,wang2005b,lo2005}. In addition to attacking the quantum channel, Eve listens on the classical channel between Alice and Bob. Afterwards Eve performs the same classical post-processing as Bob to obtain the identical secret key.

Note that the classical optical power threshold has to be sufficiently well defined for successful perfect eavesdropping. To be precise, let an optical power of $P_{100\%,i}$ or greater always cause a click when applied to detector $i$. Likewise, let an optical power of $P_{0\%,i}$ or less never cause a click when applied to detector $i$. The sufficient condition for Eve to be able to make any single detector click while none of the other detectors click, can be expressed as
\begin{equation}
  \max_i \left\{ P_{100\%,i} \right\} < 2\left(\min_i \left\{ P_{0\%,i} \right\}\right).
  \label{eq:attack-requirement}
\end{equation}

\section{Detector design}
\label{sec:detectors}
\subsection{Detector circuit}
\begin{figure}[htbp]
  \includegraphics[width=8.6cm]{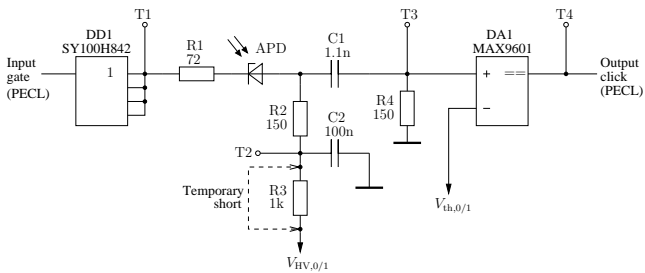}
  \caption{Equivalent detector bias and comparator circuit. Taps T1-T3 are analog taps of the APD gates ($V_{\text{gate},0/1}$), the APD bias ($V_{\text{bias},0/1}$) and the comparator input ($V_{\text{comp},0/1}$). The digital tap T4 of the detector output ($V_{\text{click},0/1}$) has been converted to logic levels in all oscillograms. For the experiments presented in section~\ref{sec:blinding}, the resistor R3 has been shorted.}
  \label{fig:detector-circuit}
\end{figure}

Figure~\ref{fig:detector-circuit} shows an equivalent detector bias and comparator circuit diagram for the detectors in Clavis2, based on reverse engineering.
The APD is biased just above its breakdown voltage by the high voltage supply $V_{\text{HV,0}} = -42.89\,\volt$, $V_{\text{HV,1}} = -43.08\,\volt$. On top of this bias the APD is gated with $2.8\,\nano\second$ TTL pulses every $200\,\nano\second$ from DD1 to create Geiger mode gates. The gates are applied as PECL signals from the mainboard, and the buffer converts them to TTL levels, $0\,\volt$ and approximately $3\,\volt$. The anode of the APD is AC-coupled to a fast comparator DA1 with the thresholds $V_{\text{th,0}} = 78\,\milli\volt$ and $V_{\text{th,1}} = 82\,\milli\volt$.

\begin{figure}[htbp]
   \centering
   \includegraphics[width=\oscwidth]{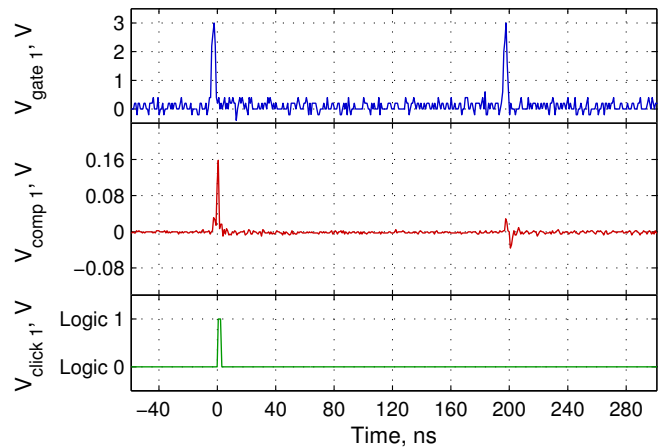}
   \caption{An example of electrical signals during two gates in detector~1 without any illumination. In the first gate thermal fluctuations or trapped carriers have caused an avalanche, and a click at the comparator output (dark count). A typical amplitude of the avalanche peak is $200\,\milli\volt$ for detector~0 and $300\,\milli\volt$ for detector~1. Normally the system removes 50 gates after a detection event, but for this oscillogram this feature has been disabled. In the second gate there is no detection event. When no current runs through the APD, it is equivalent to a capacitor, and thus approximately the derivative of the gate pulse shape propagates to the comparator input, with peak positive amplitude $\approx 35\,\milli\volt$.}
   \label{fig:D1-normal-operation-click} 
\end{figure}

The normal operation of the detector circuit can be seen in Fig.~\ref{fig:D1-normal-operation-click}. A number of techniques have been developed for compensating the capacitive pulse through APDs in the absence of an avalanche \cite{cova1981,bethune2000,tomita2002,yuan2007}, but this particular detector simply sets the comparator thresholds above the amplitude of the capacitive pulse.

As a side note, applying CW illumination to the APD allowed us to measure the timing of the quantum efficiency curve within the gate quite precisely, see Appendix~\ref{sec:gate-vs-quantum-efficiency}.

\subsection{Detector cooling}
To reduce the probability of dark counts, APDs are usually cooled to a low temperature. The two APDs in this QKD system are cooled together by one 4-stage thermoelectric cooler (TEC) (Osterm PE4-115-14-15 \cite{osterm-PE4}). The system software reports the temperature measured by a thermistor mounted on the cold side of the top stage (cold plate), and close to where the APDs are mounted. Note that the cold plate temperature is not always the same as the APD chip temperature, as there is actually a quite substantial thermal resistance between the two. This will become an important point in section~\ref{sec:frames}. The hot side of the TEC is mounted on a large heatsink with a fan, such that it stays at approximately room temperature.

The temperature of the cold plate is maintained at a pre-set value by a closed-loop controller that adjusts the TEC current. When the system is switched on, the cold plate (and thus the APDs) is first cooled to the target temperature, $-50\,\celsius$. The system will not start operation unless the cold plate settles at a temperature below $-49.8\,\celsius$. After this initial check however, during system operation, there seems to be no future checks of the cold plate temperature, even if the controller is unable to keep it at the target value.

\section{Blinding and control}
\label{sec:blinding}
Blinding is achieved when the system is insensitive to single photons. This can be achieved by ensuring that the APD bias voltage is below the breakdown voltage, or by lowering the voltage in front of the comparator such that the avalanche current does not cross the comparator threshold. The detectors are controllable if they are accessible to Eve in the linear mode with a sufficiently well defined classical optical power click threshold, as in Eq.~\ref{eq:attack-requirement}.

We have previously reported that blinding Clavis2 can be achieved by CW illumination due to the bias voltage supply impedance R3 $= 1\,\kilo\ohm$, which makes the bias voltage drop to a level where the APD is never in Geiger mode \cite{lydersen}, even inside the gate.

One fast and easy countermeasure could be to use a low-impedance bias voltage source in the detectors. Therefore, in this paper we consider a modified version of the detectors with R3 shorted (see Fig.~\ref{fig:detector-circuit}). We present three different blinding techniques which may be used against detectors with a low-impedance bias voltage source, and show that the detectors can be controlled by trigger pulses in the blind state. The technique in section~\ref{sec:thermal} clearly works against high-impedance biased detectors as well as against low-impedance biased detectors since it has been demonstrated \cite{lydersen}. The difference is that with a low-impedance bias voltage source, the blinding originates from thermal effects instead of bias voltage drop. The technique in section~\ref{sec:frames} has been used on low-impedance biased detectors, but we see no reason why it should not work similarly well against the unmodified high-impedance biased detectors. The technique in section~\ref{sec:sinkhole} has been used on both high- and low-impedance biased detectors, but we only present the results for the low-impedance biased detectors in this paper.

\subsection{Thermal CW-blinding}
\label{sec:thermal}
\begin{figure}[htbp]
  \includegraphics[width=\plotwidth]{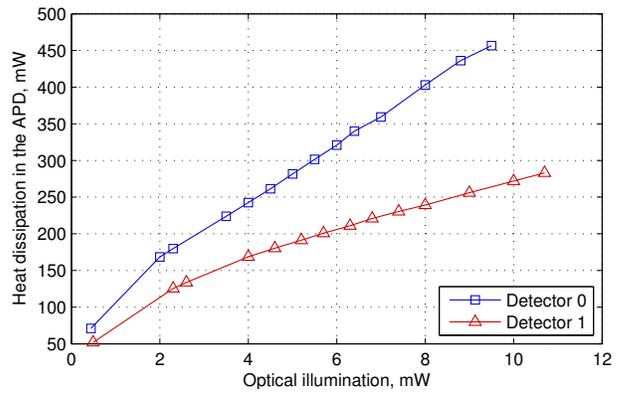}
  \caption{Calculated heat dissipation (based on measured APD current and voltage) versus the optical illumination for each of the two detectors.}
  \label{fig:plaser-vs-heat} 
\end{figure}

It turns out that it is possible to blind also low-impedance biased detectors (R3 = 0) by CW illumination. When an APD is illuminated, the power dissipated in the APD is transformed to heat, which may increase the APD temperature. The breakdown voltage is temperature dependent: increasing the temperature increases the breakdown voltage. Since the bias voltage is constant, this makes the APD leave the Geiger mode. Two effects contribute to the power dissipation: electrical heating ($V_{\text{APD}}\cdot I_{\text{APD}}$) and the small contribution by the absorption of the optical power. For the heat dissipation calculations, we simply assume that all the optical power is absorbed and transformed to heat. Figure~\ref{fig:plaser-vs-heat} shows how the heat dissipation increases with the optical illumination.

\begin{figure}[htbp]
   \includegraphics[width=\plotwidth]{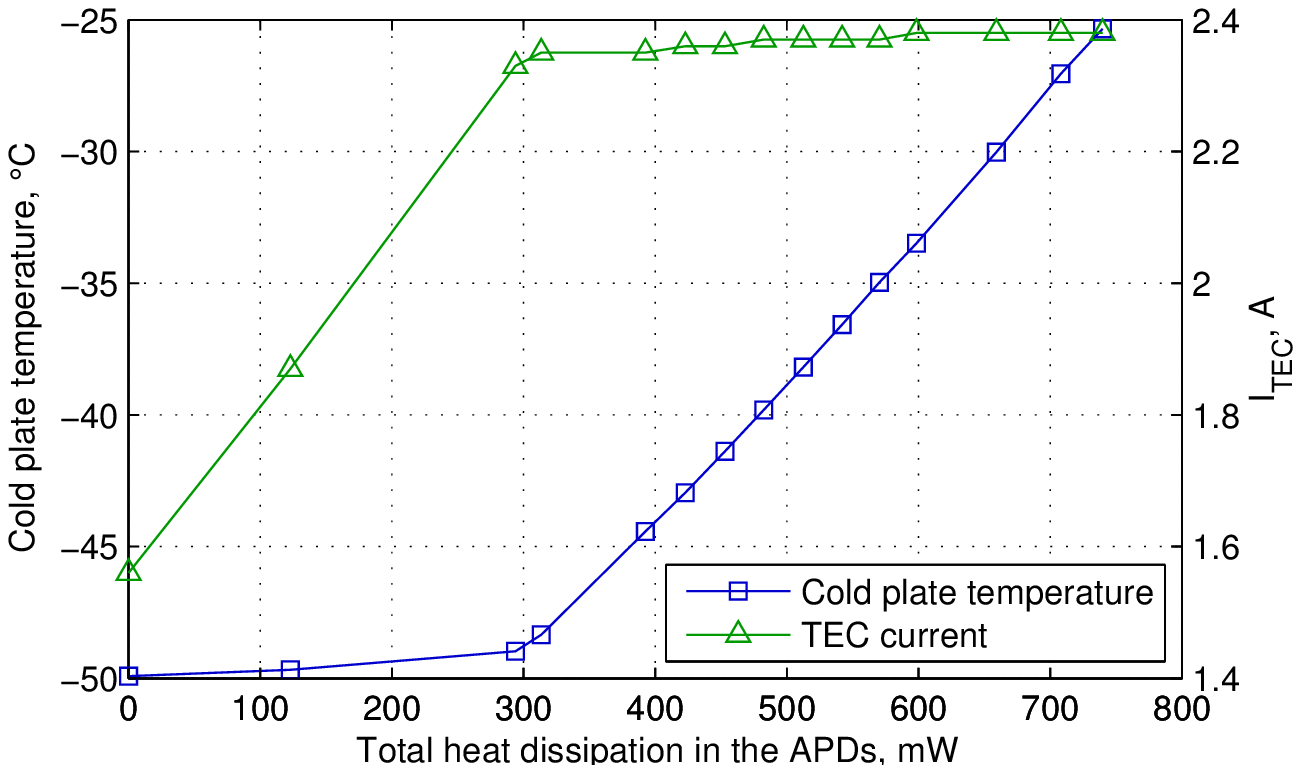}
   \caption{The temperature of the cold plate and TEC current reported by the software, versus the total amount of heat dissipated in the APDs. It takes several minutes for the cold plate temperature to stabilize at a new value (hotter than $-50\,\celsius$) after the power dissipation in the APDs is changed.}
   \label{fig:heat-vs-coldtemp} 
\end{figure}

When the sum of the heat dissipations of the two detectors is approximately $300\,\milli\watt$, the cooling system is running at its maximum capacity with a TEC current of about $I_{\text{TEC}} = 2.37\,\ampere$ (the air temperature at the heatsink fan intake at this time was $23.6\,\celsius$). When the optical illumination is increased beyond this point, the cold plate (and thus APD) temperature starts to increase. Figure~\ref{fig:heat-vs-coldtemp} shows how the temperature of the cold plate increases with the total amount of heat dissipated in the APDs. When the optical illumination, and thus the load is increased beyond the maximum capacity of the TEC, the cold plate temperature increases approximately linearly with the heat dissipated by the APD. While not in the specifications of this specific TEC \cite{osterm-PE4}, other data sheets of similar TECs \cite{marlowIndustries} show that the temperature difference between the hot and cold plate decreases linearly with respect to the load, given a constant TEC current.

When the temperature of the APDs increases, the breakdown voltage also increases with the coefficient of about $0.1\,\volt/\kelvin$ \footnote{The detectors do not have any dark counts and are assumed blind at a temperature of about $-40\,\celsius$ at the cold plate, or when the bias voltage is decreased by $0.97\,\volt$. If one assumes that the APD temperature is equal to the cold plate temperature, this means that heating the detectors by $10\,\kelvin$ is equivalent to decreasing the bias voltage by about $1\,\volt$.}. In this experiment we illuminated both detectors simultaneously, to get sufficient temperature increase without risking a permanent damage to the APDs. We used a fibre-optic coupler (see appendix~\ref{sec:setup} for the experimental setup) to illuminate both detectors, with 46.75\%/53.25\% of the optical power going to detector 0/1. This is approximately equal to the measured splitting ratio for the beam splitter in front of the detectors in the system, when illuminated through the short arm of the interferometer \cite{ribordy1998,stucki2002,gisin2002}.

\begin{figure}[htbp]
   \includegraphics[width=\plotwidth]{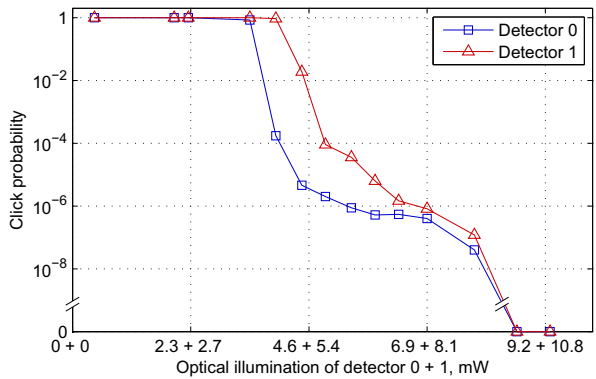}
   \caption{Click probability versus power of CW illumination applied to both detectors simultaneously.}
   \label{fig:plaser-vs-click-prob} 
\end{figure}

Fig.~\ref{fig:plaser-vs-click-prob} shows the click probability versus the CW illumination of the two detectors. The click probability drops below the normal dark count probability (about $10^{-4}$), before it becomes \emph{exactly zero} when the illumination exceeds $8.8\,\milli\watt$ and $10\,\milli\watt$ at the detectors. In the experiment the blinding caused clicks for several minutes before the APDs were properly heated. However, the blinding only needs to be turned on once, afterwards Eve remains undetected.

After the cold plate has been heated by APD illumination, it takes several tens of seconds before it cools to the target temperature of $-50\,\celsius$. Therefore, the detectors stay blind for some time after the CW blinding illumination is turned off. Detectors~0 and 1 regain dark counts when the cold plate (and thus the APDs) becomes colder than $-39.8\,\celsius$ and $-40.1\,\celsius$, respectively.

\begin{table}[htbp]
  \caption{Control pulse peak power at 0~\% and 100~\% click probability thresholds, in CW thermal blinding mode.}
  \begin{tabular}{c|cc}
    \multirow{2}{*}{Detector} &\multicolumn{2}{c}{Click probabilities}\\
      &  0 \%  &  100 \%  \\ \hline
    0 & $1.12\,\milli\watt$ & $1.31\,\milli\watt$\\
    1 & $1.71\,\milli\watt$ & $2.02\,\milli\watt$\\
  \end{tabular}
  \label{tab:highcw-det-prob-thresholds} 
\end{table}

To verify that the detectors could be controlled, the detectors were blinded with $9.5\,\milli\watt$ at detector~0 and $10.7\,\milli\watt$ at detector~1, and controlled by superimposing a $3\,\nano\second$ long laser pulse slightly after the gate. The click probability thresholds are listed in table~\ref{tab:highcw-det-prob-thresholds}. The thresholds satisfy Eq.~\ref{eq:attack-requirement}, and thus the eavesdropping method described in section~\ref{sec:attack} should be possible when the detectors are thermally blinded by CW illumination.
 
After observing thermal blinding in this experiment, we realized that this could be the reason why the PerkinElmer SPCM-AQR actively-quenched detector module remained blind at bright pulse frequencies above $400\,\kilo\hertz$, despite no substantial bias voltage drop \cite{makarov2008a}. Therefore we did more precise measurements which confirm that PerkinElmer SPCM-AQR can be thermally blinded \cite{sauge}.

\subsection{Thermal blinding of frames}
\label{sec:frames}
As this QKD system is of plug-and-play type, it sends the qubits in packets called \emph{frames} to avoid Rayleigh back-scattered photons to arrive during the gates and increase the QBER \cite{ribordy1998,ribordy2000}. For our experiment we used 1072 qubits per frame \footnote{The system actually sends the qubits in frames of 1075 qubits each. We initially made a mistake when counting them and used 1072 qubits, which is very close and does not affect the results.}. With a $200\,\nano\second$ bit period this makes the frame length $214.4\,\micro\second$. The break in between the frames varies with the fibre length between Alice and Bob, but is always longer than the frame itself. In our experiment we simply used a $250\,\micro\second$ frame break, which makes a total frame + break period of $464.4\,\micro\second$.

It turns out that the APD chip and the inner parts immediately touching it (\emph{not} the APD package and not the cold plate) act as a thermal reservoir on the frame period time scale. Therefore bright illumination between the frames heats the APD sufficiently that it stays blind throughout the whole frame. Based on the optical power where the frames went blind, and the average current through the APDs, the thermal resistance between each APD chip and the cold plate is estimated to be at least $190\,\kelvin/\watt$.

\begin{figure}[htbp]
   \includegraphics[width=\oscwidth]{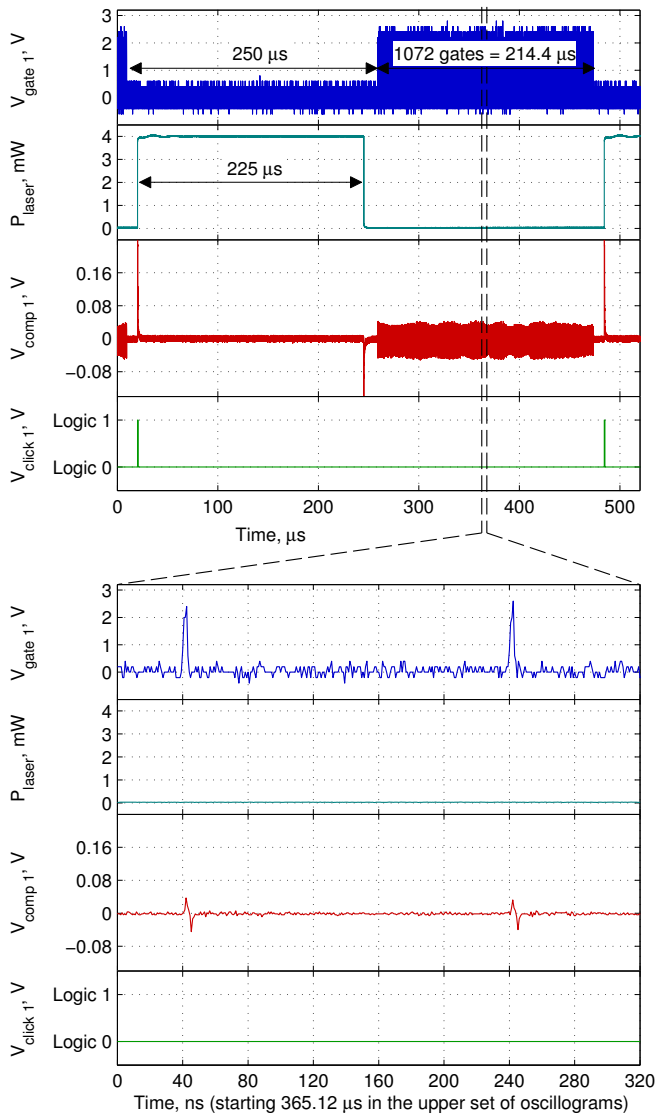}
   \caption{Thermal blinding of frames. The oscillograms show electrical and optical signals when frames of 1072 gates in detector~1 are thermally blinded by a $225\,\micro\second$ blinding pulse, with $3.5\,\milli\watt$ pulse power at detector~0, and $4\,\milli\watt$ pulse power at detector~1. The blinding pulse causes a detection event outside the frame, where the system probably does not register clicks (If the click is registered, it could easily be avoided by increasing the power of the blinding pulse gradually, such that the comparator input AC-coupling keeps the voltage below the comparator threshold).}
   \label{fig:D1-framed-blinding} 
\end{figure}

To heat the APDs we used $225\,\micro\second$ long pulses timed in between the frames and fired at both APDs simultaneously. The whole frame went blind at approximately $1.5\,\milli\watt$ and $1.7\,\milli\watt$ pulse power at detector~0 and 1 respectively. The oscillograms in Fig.~\ref{fig:D1-framed-blinding} show the electrical and optical signals in detector~1 when frames of 1072 gates are thermally blinded by the $225\,\micro\second$ long pulses with $3.5\,\milli\watt$ in-pulse power at detector~0, and $4\,\milli\watt$ in-pulse power at detector~1.
While the system was blind, the cold plate temperature reading was $-49.5\,\celsius$, and the TEC was running well below its maximum capacity at $I_{\text{TEC}} = 2.006\,\ampere$. 

\begin{figure}[htbp]
  \includegraphics[width=\oscwidth]{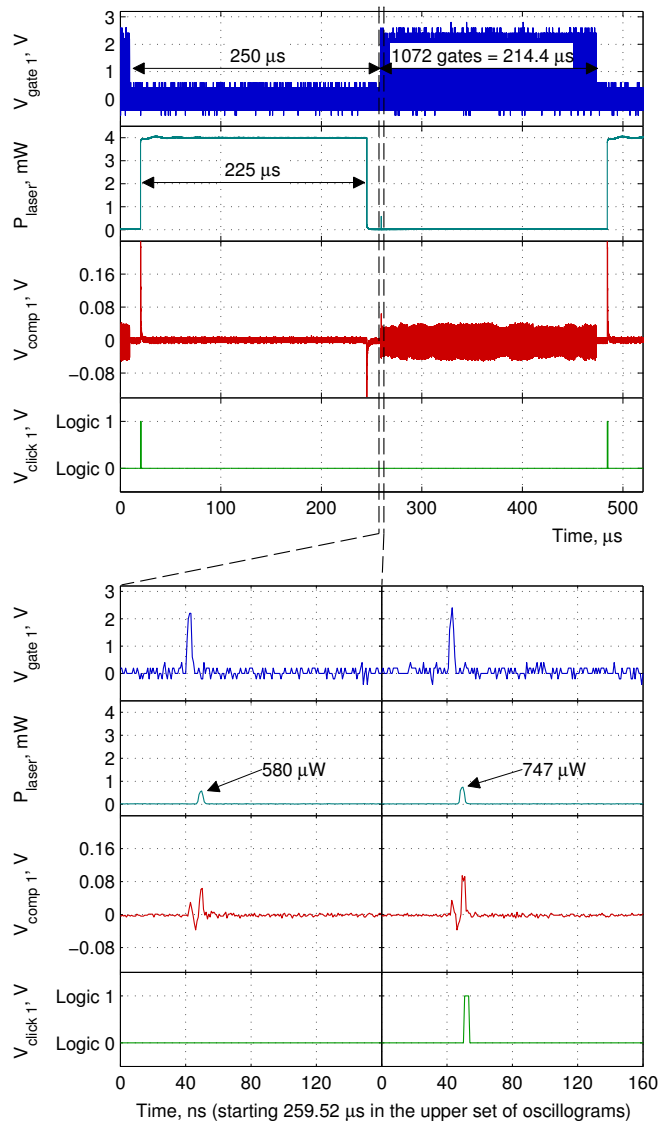}
  \caption{Detector control during thermal blinding of frames. The oscillograms show electrical and optical signals when frames of 1072 gates in detector~1 are thermally blinded by a $225\,\micro\second$ blinding pulse, with $3.5\,\milli\watt$ pulse power at detector~0, and $4\,\milli\watt$ pulse power at detector~1, and the detector is controlled by a $4\,\nano\second$ long control pulse timed slightly after the second gate in the frame. In the upper and lower left sets of oscillograms, the $580\,\micro\watt$ control pulse never causes any click. In the lower right set, the control pulse is applied after the same gate in the frame, but now its increased $747\,\micro\watt$ peak power always causes a click.}
  \label{fig:D1-frames-blinding-control-last} 
\end{figure}

To verify that the detectors could be controlled, we checked the response to a $4\,\nano\second$ long control pulse timed slightly after the gate of one of the first bits of the frame, and the last bit of the frame. The detection probability thresholds for the second \footnote{We picked the second bit to simplify synchronization in our measurement setup. The results for the first bit should be very similar to the results for the second bit.} and the last bit are given in tables~\ref{tab:frames-det-prob-thresholds-first} and \ref{tab:frames-det-prob-thresholds-last}. Figure~\ref{fig:D1-frames-blinding-control-last} shows oscillograms from detector~1 when it is blinded and controlled in the second bit of the frame.

The click probability thresholds in tables~\ref{tab:frames-det-prob-thresholds-first} and \ref{tab:frames-det-prob-thresholds-last} each satisfy Eq.~\ref{eq:attack-requirement} individually. However, $P_{0\%,0}$ in the last bit of the frame is less than $1/2$ of $P_{100\%,1}$ in the second bit of the frame. This means that the control pulse power would have to be decreased throughout the frame. Since the second and the last bit of the frame can be controlled, it is plausible that the eavesdropping method described in section~\ref{sec:attack} could be applied to any bit of the frame.

What is remarkable about this blinding method is that due to the low thermal conductivity between the APD chip and the cold plate, as well as the thermal inertia of the nearby parts, the cold plate thermistor reports a value very close to the normal value. Therefore monitoring the cold plate temperature would not suffice to prevent thermal blinding.

\begin{table}[htbp]
  \caption{Control pulse peak power at 0~\% and 100~\% click probability thresholds for the second bit in the frame, when the frame is thermally blinded.}
  \begin{tabular}{c|cc}
    \multirow{2}{*}{Detector} &\multicolumn{2}{c}{Click probabilities}\\
      &  0 \%  &  100 \%  \\ \hline
    0 & $401\,\micro\watt$ & $533\,\micro\watt$ \\
    1 & $580\,\micro\watt$ & $747\,\micro\watt$ \\
  \end{tabular}
  \label{tab:frames-det-prob-thresholds-first} 
\end{table}
\begin{table}[htbp]
  \caption{Control pulse peak power at 0~\% and 100~\% click probability thresholds for the last bit in the frame, when the frame is thermally blinded.}
  \begin{tabular}{c|cc}
    \multirow{2}{*}{Detector} &\multicolumn{2}{c}{Click probabilities}\\
      &  0 \%  &  100 \%  \\ \hline
    0 & $305\,\micro\watt$ & $420\,\micro\watt$ \\
    1 & $340\,\micro\watt$ & $532\,\micro\watt$ \\
  \end{tabular}
  \label{tab:frames-det-prob-thresholds-last} 
\end{table}

\subsection{Sinkhole blinding}
\label{sec:sinkhole}
It is natural to ask whether the framed blinding technique can be applied at the single gate level, i.e. what happens if bright illumination is applied between adjacent gates? It turns out that this also leads to blinding, but not primarily due to thermal effects. Since the comparator input is AC-coupled (see Fig.~\ref{fig:detector-circuit}), the signal at the input of the comparator has the same area over and under $0\,\volt$ level when averaged over time much longer than R4$\cdot$C1 = $165\,\nano\second$. Thus by sending long bright pulses between the gates and no illumination near the gate, it is possible to superimpose a negative-voltage pulse at the comparator input at the gate time. We call this negative pulse a \emph{sinkhole}. An avalanche that occurs within it can have a normal amplitude yet remain below the comparator threshold level.

\begin{figure}[htbp]
  \includegraphics[width=\oscwidth]{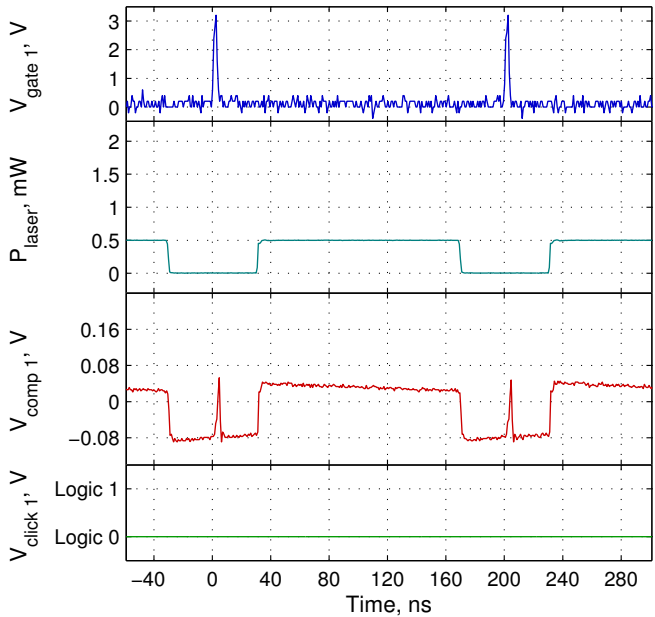}
  \caption{Sinkhole blinding. The oscillograms show electrical and optical signals when detector~1 is blinded by a $500\,\micro\watt$, $140\,\nano\second$ long laser pulse in between the gates. The avalanche amplitude is about $130\,\milli\volt$ and would cause a click if it were not sitting in the negative-voltage pulse. It seems that the reduction in avalanche amplitude (compare to Fig.~\ref{fig:D1-normal-operation-click}) is caused by heating of the APD, which effectively rises the breakdown voltage.}
  \label{fig:D1-sinkhole-blind}
\end{figure}

\begin{table}[bhtp]
  \caption{Control pulse peak power at 0~\% and 100~\% click probability thresholds, during sinkhole blinding.} 
  \begin{tabular}{c|cc}
    \multirow{2}{*}{Detector} &\multicolumn{2}{c}{Click probabilities}\\
      &  0 \%  &  100 \%  \\ \hline
      0 & $655\,\micro\watt$ & $751\,\micro\watt$ \\
      1 & $773\,\micro\watt$ & $908\,\micro\watt$ \\
  \end{tabular}
  \label{tab:sinkhole-control-thresholds}
\end{table}

Using a $140\,\nano\second$ long pulse beginning about $25\,\nano\second$ after the gate, detector~0 becomes completely blind when $P_{\text{laser}} > 205\,\micro\watt$, and detector~1 becomes blind when $P_{\text{laser}} > 400\,\micro\watt$. To keep both detectors blind, $P_{\text{laser}} = 500\,\micro\watt$ is used subsequently. When a large pulse is applied between the gates, the detector will always experience a dark count in the gate due to trapped carriers. Figure~\ref{fig:D1-sinkhole-blind} shows detector~1 blinded by a $140\,\nano\second$ long, $500\,\micro\watt$ bright pulse, starting about $25\,\nano\second$ after the gate.

Initially when the blinding pulses are turned on, there is a transient with about 20-100 clicks, which would be easily detectable in post-processing. Note again that the blinding only needs to be turned on once, and that the blinding can be turned on before the raw key exchange to avoid the clicks being registered.

\begin{figure}[thbp]
  \includegraphics[width=\oscwidth]{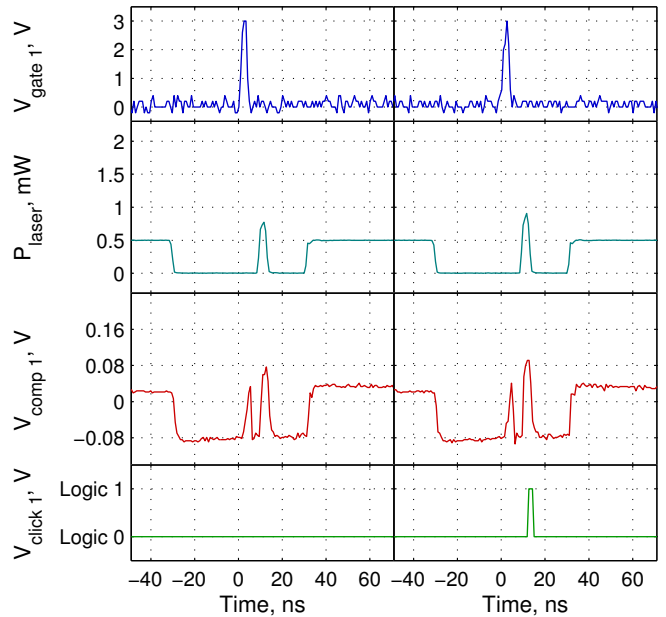}
  \caption{Detector control during sinkhole blinding. The oscillograms show electrical and optical signals when detector~1 is blinded with a $500\,\micro\watt$, $140\,\nano\second$ long laser pulse in between the gates, and controlled with a $3.2\,\nano\second$ long laser pulse timed shortly after the gate. To the left, the $773\,\micro\watt$ control pulse never causes any click. To the right, the $908\,\micro\watt$ control pulse always causes a click.}
  \label{fig:D1-sinkhole-blinding-and-control}
\end{figure}

Detector control is obtained by a $3.2\,\nano\second$ long laser pulse timed shortly after the gate. The click probability thresholds found are listed in Table~\ref{tab:sinkhole-control-thresholds}. Figure~\ref{fig:D1-sinkhole-blinding-and-control} shows oscillograms from detector~1 when it is blind and controlled. 
Once again, the thresholds in table~\ref{tab:sinkhole-control-thresholds} satisfy Eq.~\ref{eq:attack-requirement}, and thus the eavesdropping method described in section~\ref{sec:attack} should be possible when the detectors are sinkhole blinded.

\section{Discussion and countermeasures}
\label{sec:countermeasures}
First of all, the numerous detectors proved blindable and controllable \cite{makarov2009,makarov2008a,sauge,lydersen,gerhardt}, and the large number of independent blinding methods available show that avoiding this loophole is non-trivial. Further the results presented in this paper clearly show that reducing the impedance of the bias voltage supply is far from being a sufficient countermeasure for this detector design.

At this point it is not clear to us how to design hack-proof detectors. The most obvious countermeasure is to monitor the optical power at Bob's entrance with an additional detector. However it is not obvious that this actually closes the loophole; as pointed out previously the click threshold close to the gate may be very low, allowing for practically non-detectable control pulses \cite{lydersen}. Thus it is not clear how to set the threshold value for the entrance monitor; in any case the threshold should be derived from and incorporated into a security proof. It would also be crucial that this monitoring detector is not blindable.

For the passively quenched scheme it has been proposed previously to monitor APD parameters such as APD bias voltage, current and temperature \cite{makarov2009}. However, the results in section~\ref{sec:frames} show that normal APD parameters do not necessarily guarantee single photon sensitivity: for thermal blinding of frames all the APD parameters report normal values during the frames while the detectors are in fact blind.

It is worth emphasizing that the loophole opens when Eve drives the detectors into an abnormal operating regime, namely the linear mode. However, there are also quantum detectors which are actually designed to operate in linear mode.  For example, homodyne detectors used in continuous-variable QKD \cite{braunstein2005,andersen2010} are probably not susceptible to the described attack.

\section{Conclusion}
\label{sec:conclusion}
The detectors in the Clavis2 QKD system have proved to be blindable by a variety of methods, even with a low-impedance bias voltage supply. Further, the detectors can always be controlled in the blind state. This allows eavesdropping on the QKD system, using the method described in section~\ref{sec:attack}. Since Eve may use an exact copy of Bob's system, no parameters currently available to Bob reveal Eve's presence. In practice, this should allow for perfect eavesdropping where Eve has an exact copy of Bob's raw key, and thus can extract the full secret key. The eavesdropping strategy described in section~\ref{sec:attack} has been implemented and used to capture 99.8\% of the raw key in a $290\,\meter$ experimental entanglement-based QKD system \cite{gerhardt}. We see no practical difficulties implementing the same eavesdropper for this commercial QKD system, using off-the-shelf components. Actually we have proposed a plug-and-play eavesdropper scheme \cite{lydersen} for easy deployment.

Many detectors have already been proved blindable and controllable by Eve \cite{makarov2009,makarov2008a,lydersen}, and the large variety of blinding methods available for the system tested could probably be used on other detector designs as well. While it is relatively easy to design a countermeasure that prevents blinding attacks with the specific parameters chosen in the present work, it is unclear to us how to build generic secure detectors.

This work further emphasizes the importance of thoroughly investigating the non-idealities of each component in a QKD system, as well as battle-testing the system as a whole.

ID Quantique has been notified about the loophole prior to this publication, and has implemented countermeasures.

\appendix
\section{Measurement setup}
\label{sec:setup}
\begin{figure}[htbp]
   \includegraphics[width=8.6cm]{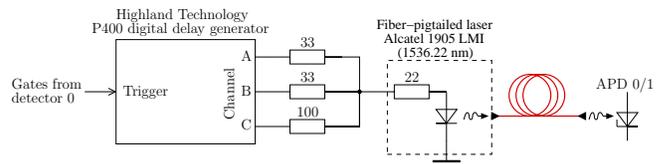}
   \caption{The setup used in the experiment. Both detectors were illuminated simultaneously by inserting a 50/50 fibre-optic coupler (not shown in the diagram) before the APDs.}
   \label{fig:setup} 
\end{figure}

Figure~\ref{fig:setup} shows the measurement setup used for this experiment. The trigger signal is tapped directly from the PECL gate signal (before DD1 in Fig.~\ref{fig:detector-circuit}).

When pump current is used to control the power of the laser, the pulse width will vary slightly with the peak power. In our experiment, the observed change in pulse width is less than 10 \% after doubling the laser power. Also, the comparator threshold does not seem to be significantly dependent on the pulse width, thus we consider our results valid despite this small change in the laser pulse width.

\section{Direct measurement of quantum efficiency}
\label{sec:gate-vs-quantum-efficiency}
When CW illumination is applied to the APD, the applied electrical gate ``propagates'' to the comparator input. This might be caused by a change in linear multiplication coefficient caused by the electrical gate. This allowed us to measure the quantum efficiency mapped inside the ``propagated'' gate with about $200\,\pico\second$ precision.

\begin{figure}[ht!]
   \includegraphics[width=\plotwidth]{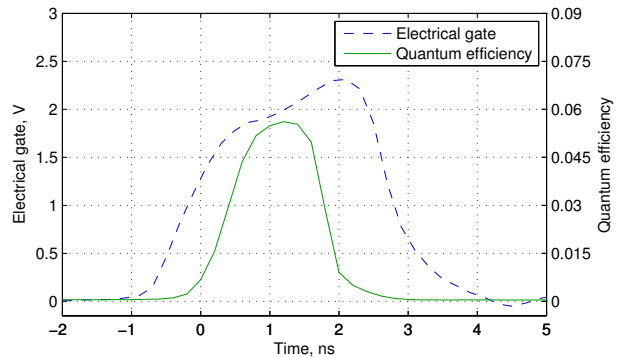}
   \caption{Quantum efficiency measured directly within the electrical gate for detector~1. The photon sensitivity drops about $1\,\nano\second$ before the falling edge of the gate, because avalanches that start late do not have time to develop a large enough current to cross the comparator threshold.}
   \label{fig:gate-vs-quantum-efficiency}
\end{figure}

The single photon sensitivity was measured using a id300 short-pulsed laser attenuated to a mean photon number of 1 per pulse. The quantum efficiency $\eta$ was derived from the data assuming that the detector is linear (i.e. that an n-photon state is detected with probability $1 - (1 - \eta)^n$). The timing of the photon arrival at the APD relative to the applied gate was aligned by observing a response to unattenuated laser pulse on top of the $2.1\,\milli\watt$ CW illumination. Figure~\ref{fig:gate-vs-quantum-efficiency} shows the result of the measurement on detector~1.

\begin{acknowledgments}
This work was supported by the Research Council of Norway (grant no{.} 180439/V30) and DAADppp mobility program financed by NFR (project no{.} 199854) and DAAD (project no{.} 50727598).
\end{acknowledgments}

%\bibliography{bibtex_library}% Produces the bibliography via BibTeX.
%merlin.mbs 2010-03-15 4.21a (PWD, AO, DPC)

\end{document}